\documentstyle[preprint,aps]{revtex}  

\newcommand{\ul}[1]{\underline{#1}}
\newcommand{\E}{{\cal E}}
\tighten       

\begin{document}
\draft

\title{Microscopic modelling of perpendicular electronic transport in doped 
multiple quantum wells}
\author{Andreas Wacker and Antti-Pekka Jauho}
\address{Mikroelektronik Centret, Danmarks Tekniske Universitet, 
DK-2800 Lyngby}
\date{to be published in Physica Scripta, Proceedings of the $17^{{\rm th}}$
Nordic Semiconductor Meeting, Trondheim, June 1996}
\maketitle

\begin{abstract}
We present a microscopic calculation of
transport in strongly doped superlattices
where domain formation is likely to occur.
Our theoretical method is based on a current formula involving
the spectral functions of the system, and thus allows,
in principle, a systematic investigation of various
interaction mechanisms.
Taking into account impurity scattering and optical 
phonons we obtain a
good quantitative agreement with existing experimental data from
Helgesen and Finstad (J.~Appl.~Phys. {\bf 69}, 2689, (1991)).
Furthermore the calculated spectral functions indicate a significant
increase of  the average intersubband spacing compared to the bare 
level differences which might explain the experimental trend.
\end{abstract}

\section{Introduction}
In semiconductor superlattices the electric transport is dominated
by resonances between the localized energy levels inside the wells.
A resonance structure appears in the current-voltage characteristics
if the energy levels of different wells align\cite{KAZ72,ESA74,CAP86}.
Between these resonances regions with negative differential 
conductivity (NDC) are likely to appear.
This yields interesting current-voltage characteristics as already 
shown by Esaki and Chang \cite{ESA74}. 
If the voltage applied along the growth direction of the superlattice
yields an average field which is in the NDC region,
the homogeneous field distribution breaks up and electric
field domains form causing many current branches in the
current-voltage characteristic (almost equal to the number of quantum wells).
This has been extensively studied
experimentally during the last decade 
\cite{KAW86,CHO87,HEL90,GRA91,KAS94,KWO95}.

Theoretically the complicated measured $I(U)$ 
characteristics could be reproduced
using models based on the  combination of rate equations between
the wells and Poisson's equation \cite{PRE94,BON94,MIL94}.
These models used simplified expressions for the 
tunnelling current between the wells \cite{PRE94}, or 
a fitted local current-field characteristics \cite{BON94}.
A theoretical approach for the transport
in superlattices with strong coupling between the wells
has been reported in Ref.~\cite{LAI93} with the restriction to one 
miniband. This restriction is not appropriate for 
situations where high field domains occur, as they are close  
to the resonance between the
first and the second subband \cite{KWO95}.
Nevertheless, no quantitative calculations have been performed so far.

The aforementioned theories for domain formation showed that 
the electric field as well as the carrier density
are almost constant within the field domains. 
Thus the current density is  determined
by intrinsic features of the superlattice. 
This is a much simpler situation than in many other
semiconductor structures, such as the double-barrier resonant-tunnelling 
diode, where the contacts
strongly influence the field profile and the current densities
by carrier injection.
Furthermore the single branches extend up to a current density, where
the low field domain reaches the maximum of the current-field relation.
Therefore the current maxima of the branches are a good estimate
for the maximum in the current-field relation, which does not
depend strongly on the contacts.
The situation becomes slightly more complicated, if the superlattice
exhibits fluctuation in its parameters, which have a strong
impact impact on the individual branches \cite{WAC95c}.

In this paper we develop a unified 
microscopic theory, without adjustable parameters,
to describe the underlying current-field relation.
In contrast to the situation considered in Ref.~\cite{LAI93},
we consider the situation of weakly coupled quantum wells ($H_1\ll \Gamma$)
where $H_1$ is the coupling between the wells and $\Gamma$ is the
broadening of the levels within the single wells.
This means that the intrawell scattering rate is 
larger than the tunnelling rate.  
This is relevant to most experiments including that 
described in Ref.~\cite{HEL90} with which we will compare our results.

\section{The Model}
We consider weakly coupled semiconductor quantum wells of period $d$. 
Then the electrons are localized in the wells and
a reasonable basis set of wave functions is given by
a product of  Wannier functions $\Psi^{\nu}(z-nd)$
localized in well $n$, and plane waves $e^{i\ul{k}\cdot \ul{r}}$.
Here the $z$ direction is defined to be the growth direction and
$\ul{k},\ul{r}$ are vectors within the $(x,y)$ plane.
$\nu$ denotes the subband within the well.

Restricting  ourselves to the lowest two minibands (denoted by $a$ and $b$) 
and coupling between neighbouring wells we consider the following hamiltonian
($F$ is the electric field, and $e<0$ is the charge of the electron): 
\begin{eqnarray}
\lefteqn{\hat{H}_0=\sum_{n,\ul{k}} \left[
E_n^{a}(\ul{k})a_n^{\dag}(\ul{k})a_n(\ul{k})
+E_n^{b}(\ul{k})b_n^{\dag}(\ul{k})b_n(\ul{k})
\right]}\label{Eqham1}\\
\lefteqn{\hat{H}_1=\sum_{n,\ul{k}} \left[
T_1^a a_{n+1}^{\dag}(\ul{k})a_n(\ul{k})
+T_1^b b_{n+1}^+(\ul{k})b_n(\ul{k})\right.}\nonumber \\
&&\left. -eFR^{ab}_1 a_{n+1}^{\dag}(\ul{k})b_n(\ul{k})
-eFR^{ba}_1 b_{n+1}^{\dag}(\ul{k})a_n(\ul{k}) 
+h.c.\right]\label{Eqham2}\\
\lefteqn{\hat{H}_2=\sum_{n,\ul{k}}
\left[ -eF(R_0^{ab}a_n^{\dag}(\ul{k})b_n(\ul{k})
+R_0^{ba}b_n^{\dag}(\ul{k})a_n(\ul{k}))\right]}
\label{Eqham3}
\end{eqnarray}
with $E_n^{\nu}(\ul{k})=E^{\nu}+\hbar^2k^2/(2m_w)-eFn$ 
($m_w$ is  the effective mass  in the well), and  the couplings 
$R_h^{\nu'\nu}=\int dz \Psi^{\nu'}(z-hd)z\Psi^{\nu}(z)$.
$4|T_1^a|$ and $4|T_1^b|$ are the miniband widths of subband $a$ and $b$,
respectively. 
The term $\hat{H}_2$ can be incorporated into the one electron states 
by diagonalizing $\hat{H}_0+\hat{H}_2$ \cite{KAZ72}. 
This leads to renormalized
coefficients in $\hat{H}_0$ and $\hat{H}_1$ but does not change the
structure of the problem for a homogeneous electric field.
In a similar way higher subbands and continuum states will change
the coefficients, so we prefer to omit
$\hat{H}_2$ in the following.

The total current from subband $\nu$ in well $n$ to subband 
$\mu$ in well $n+1$
can be described by the following expression which is derived in
Sect. 9.3 of Ref.~\cite{MAH90}:
\begin{eqnarray}
\lefteqn{J_{n\to n+1}^{\nu \to \mu}=2e\sum_{\ul{k}',\ul{k}}
|H_{(n+1)\ul{k},n\ul{k}'}^{\mu,\nu}|^2
\int_{-\infty}^{\infty} \frac{d\E}{2\pi \hbar}
A_n^{\nu}(\ul{k}',\E)}\cdot \label{EqJ} \\
&&\cdot A_{n+1}^{\mu}(\ul{k},\E+\mu_n-\mu_{n+1})
\left[n_F(\E)-n_F(\E+\mu_n-\mu_{n+1})\right]\, .
\nonumber 
\end{eqnarray}
Here $\mu_n$ is the electro-chemical potential in well $n$ and 
$n_F(\E)=(1+e^{\beta\E})^{-1}$ is the Fermi function.
The energy $\E$ is measured with respect to $\mu_n$.
Note that for equal densities
in both quantum wells we have $\mu_n-\mu_{n+1}=eFd$. 

$A_n^{\nu}(\ul{k},\E)$ denotes the spectral function for the state
$\ul{k}$ of the subband $\nu$ in well number $n$.
It is calculated in equilibrium neglecting the coupling to the
other wells and is related to the retarded self-energy 
$\Sigma_{n}^{\nu\,{\rm ret}}(\ul{k},\E)$ via 
\begin{equation}
A_n^{\nu}(\ul{k},\E)=\frac{-2 {\rm Im}\Sigma_{n}^{\nu\,{\rm ret}}}
{\left(\E+\mu_n- E_n^{\nu}(\ul{k})-
{\rm Re}\Sigma_{n}^{\nu\,{\rm ret}}\right)^2
+\left({\rm Im}\Sigma_{n}^{\nu\,{\rm ret}}\right)^2}
\end{equation}
If no perturbation is present the state $\ul{k}$ has a fixed energy,
$E_n^{\nu}(\ul{k})$, and the spectral function becomes
a $\delta$-function $A_n^{\nu}(\ul{k},\E)=2\pi 
\delta (\E+\mu_n-E_n^{\nu}(\ul{k}))$.
If scattering is present the states $\ul{k}$ are no longer eigenstates
of the full hamiltonian and the spectral function becomes 
smeared out, which is often 
modeled by a Lorentzian using a constant $\Sigma_{n}^{\nu\,{\rm ret}}$.

While  the full derivation is slightly tedious \cite{MAH90}
the formula (\ref{EqJ}) can be motivated quite easily:
In the long-time limit energy has to be conserved during transitions
caused by the time-independent interwell couplings 
$H_{(n+1)\ul{k},n\ul{k}'}^{\mu,\nu}$.
Therefore we have to consider tunnelling processes for
a certain energy $\E$ 
and integrate over $\E$ afterwards.
The factor $\left[n_F(\E)-n_F(\E+\mu_n-\mu_{n+1})\right]$ takes
into account the thermal occupation at the given energy in both wells.
The free particle state $\ul{k}'$ has a weight
$A_n^{\nu}(\ul{k}',\E)/(2\pi)$ in well $n$. 
Its transition probability to the state $\ul{k}$ in well $n+1$ is
given by  
$2\pi |H_{(n+1)\ul{k},n\ul{k}'}^{\mu,\nu}|^2/\hbar $. The final state
has a weight $A_{n+1}^{\mu}(\ul{k},\E+\mu_n-\mu_{n+1})/(2\pi)$ 
at the given energy.
Obviously one has to sum over all free particle states 
$\ul{k},\ul{k}'$.
Finally, the factor 2 is due to the spin degeneracy.

\begin{figure}
\vspace*{5.1cm}
\includegraphics{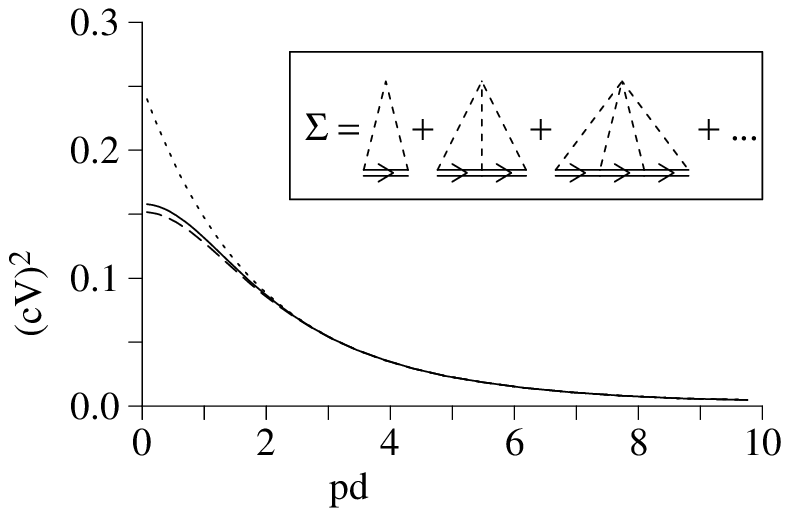}
\caption[a]{Square of the screened matrix element for impurity scattering. 
The full line gives
$\sum_h \left| V^{{\rm sc}}_h(\ul{p})\right|^2$ which determines the 
scattering in the  BA. The dashed line gives
$\left| V^{{\rm sc}}_0(\ul{p})\right|^2$. The dotted line gives the
same expression within the restriction to screening by electrons from
the same well. The inset depicts the diagrams contained in the
SSA.}
\label{Figcvq}
\end{figure}

We calculate the self-energy $\Sigma^{a\,{\rm ret}}(\ul{k},\E)$
for the lower subband for impurity scattering against the ionized donors
within the self-consistent single-site-approximation  (SSA)
(which contains all noncrossing diagrams as shown in the inset
of Fig.~\ref{Figcvq}).
This approximation was also used in Ref.~\cite{SER89} to calculate
the spectral functions in a single quantum well. 
The matrix element $V_h(\ul{p})$ of the interaction is
calculated from the given Wannier functions both for
impurities in the same well ($h=0$) and for 
remote impurities located in the $h$-th well counted from the electron,
which become important for $pdh<1$.
We consider screening by the free 2D electron gas located both in the same well
and  in different wells
within the $T=0$ Random Phase Approximation.
The screened matrix element for impurity scattering is denoted by 
$V^{{\rm sc}}_h(\ul{p})$.
Within the Born approximation the total
scattering rate with momentum transfer $\ul{p}$ is proportional to 
$\sum_h \left| V^{{\rm sc}}_h(\ul{p})\right|^2$.
In Fig.~\ref{Figcvq} we have plotted this quantity as well as
$\left| V^{{\rm sc}}_0(\ul{p})\right|^2$ as a function 
of $p$.
For comparison we have also shown the result for the case
when screening and scattering is restricted within the same well;
this is larger than $\left| V^{{\rm sc}}_0(\ul{p})\right|^2$
because of  the weaker screening.
From Fig.~\ref{Figcvq} we can see 
that even for $p=0$ the main contribution to the total scattering
$\sum_h \left| V^{{\rm sc}}_h(\ul{p})\right|^2$
comes from the scattering within the well 
$\left| V^{{\rm sc}}_0(\ul{p})\right|^2$. 
Thus we can restrict ourselves to this contribution in our case.

The self-consistent Born-Approximation (BA) (consisting of the first diagram
shown in the inset of Fig.~\ref{Figcvq}) breaks down for
matrix elements $\pi cV\approx 1$, where $c=m_w/(2\pi \hbar^2)$ is
the density of states per spin of the free 2D electron gas. 
Fig.~\ref{Figcvq} tells us that we are in the range where the
higher order diagrams become important which invokes the need for the 
SSA.

For the calculation of the self-energy 
$\Sigma^{b\,{\rm ret}}(\ul{k},\E)$ of the upper subband
we additionally include the emission of optical phonons for scattering
between the  upper to the lower subband describing
the intersubband relaxation.
This leads to an additional imaginary part $-\Gamma_{{\rm ph}}^b$
in the self-energy $\Sigma^{b\,{\rm ret}}(\ul{k},\E)$.

\section{Results of the calculation}
Here we apply our approach to the experimental  situation
of Refs.~\cite{HEL90,HEL91}. The Al$_{0.3}$Ga$_{0.7}$As  superlattice 
used there had a nominal barrier width $b=12$ nm and a
well width $w=8$ nm. The middle 7 nm of the wells were
n-doped with a doping density $N_D/A=8.75\cdot 10^{15}$/m$^2$.
We calculate the coefficients in Eqs.~(\ref{Eqham1},\ref{Eqham2},\ref{Eqham3})
within the Kronig-Penney model assuming a parabolic dispersion with
the conduction band offset $\Delta E_c=0.24$ eV,
and the effectives masses $m_w=0.067m_e$  and $m_b=0.0919 m_e$ 
for well and barrier, respectively \cite{ADA93}. 
The resulting parameters are given in Table~\ref{Thelgpar}.
Just as in the more  sophisticated calculation used in 
Ref.~\cite{HEL90,HEL91}
the calculated   subband spacing $E_b-E_a=117.9$ meV is smaller
than the experimental value 
124 meV  determined by intersubband absorption.

\begin{figure}
\vspace*{5.1cm}
\includegraphics{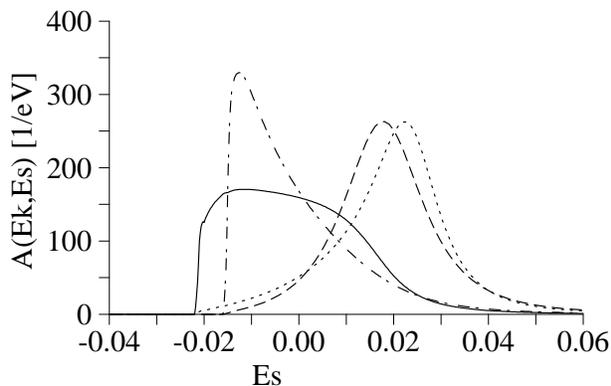}
\caption[a]{Spectral functions $A^a(E_k,E_s)$ for the first subband.
The full and dotted  line give the result within the SSA 
for $E_k=0$ and $E_k=0.02$ eV, respectively.
The dashed-dotted and dashed line give the result within the BA 
for $E_k=0$ and $E_k=0.02$ eV, respectively.
The Energy $E_s$ is measured with respect to the bottom of
the first subband.} 
\label{Figspek}
\end{figure}

We approximated the doping profile by two 
$\delta$-doping layers located at a distance of 1.75 nm from
the middle of the well.
The calculated spectral functions for subband $a$ are given in 
Fig.~\ref{Figspek} for both the BA
and the SSA.
For $E_k=0.02$ eV both spectral functions $A^{a}(E_k,E_s)$ 
(where $E_s=\E+\mu-E_{a}$ is measured with respect to $E_{a}$)
are quite similar and  have
approximately a Lorentzian shape with a half width
of $\Gamma^a\approx 0.01$ eV. In contrast to this for
$E_k=0$ the spectral function exhibit significant  differences
for the BA and the SSA. Within 
both approximations they  exhibit a sharp onset at
$E_s \approx -0.02$ eV and are significantly different from Lorentzians.

The broad width of the spectral functions 
has obviously an impact for the optical absorption $\alpha(\omega)$
between the subbands which is proportional to
\begin{equation}
\alpha(\omega)\propto \sum_{\ul{k}}\int_{-\infty}^{\infty}d\E n_F(\E) 
A^{a}(\ul{k},\E)A^{b}(\ul{k},\E-\hbar \omega)
\, .
\end{equation}
Our calculation reveals a maximum at $\hbar \omega=123.5$ meV which is
in excellent agreement with the experimental value.
Nevertheless the full width at half maximum (35 meV) is
larger than the experimental value 19 meV. 
(Within the BA the maximum is at $\hbar \omega=123$ meV and the width is
30 meV.) This shows that the measured interssubband spacing may deviate
significantly form the bare energy levels for strongly doped samples.

For the given density of carriers provided by the doping we have 
calculated the chemical potential for the actual density of
states determined from the spectral functions.

\begin{figure}
\vspace*{5.1cm}
\includegraphics{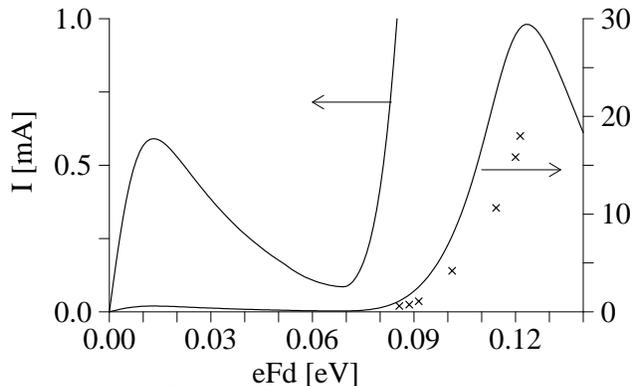}
\caption[a]{Current voltage characteristic (full line) for the parameter used
in Ref.~\cite{HEL91} except for $b=10.5$ nm. The data points (crosses)
are taken from Fig. 1 of\cite{HEL91}. Note that two different scales
at the current axis are used here for the same curve.}
\label{Figstroma}
\end{figure}

Finally we compute the current
from Eq.~(\ref{EqJ}). We use zero temperature and the
Fermi functions $n_F(\E)$ become step functions $\Theta(-\E)$.
Then only the lowest subband is occupied in equilibrium and 
the total current is given by the sum $I=I^{a\to a}+I^{a\to b}$.
Using the nominal sample parameters
we obtain current maxima  for the current-field characteristics
which are almost an order of magnitude lower than the 
currents measured in the respective regions.
This indicates that the coupling between the wells should be stronger.
Using the barrier width $b=10.5$ nm  we find good agreement with 
the experimental data. In their own  theoretical analysis of the data
Helgesen, Finstad, and Johannessen \cite{HEL91} use
the width $b=10.8$ nm, which indicates that they consider this
value to be within the experimental uncertainty.
The result of the calculation for $b=10.5$ is given 
in  Fig.~\ref{Figstroma}. The $I(eFd)$ relation exhibits
a first maximum at $eF_Md=13$ meV with a maximum current of
$I_M=0.592$ mA and a minimum at $eF_md=69$ meV 
with a current $I_m=0.086$ mA.  
At $eFd=123$ meV there is a second
maximum due to the resonance between the first and the second subband 
with a maximum current $I=29.4$ mA.
The presence of a range with negative differential conductivity for
13 meV$<eFd<$69 meV causes an instability leading to the formation
of field domains  which is discussed
in detail in Ref.~\cite{WACp} for an arbitrary $v(F)$ relation.
The domain branches  in the characteristic should exhibit
a maximum current $I_M$ if the high field domain is located at the anode
and a lower current if the high field domain is located at the cathode,
which concerns the experimental situation here\cite{HEL90}.
We obtain very similar results for the currents within the BA.

Let us compare these data with the experimental results.
In Fig. 2 of Ref.~\cite{HEL90} one can identify the first resonant
maximum very well before the domain formation sets in.
It is located at $U=0.3$ V, $I=0.64$ mA. Dividing the voltage by the number
of periods $N=23$ we obtain $eFd= 13$ meV. Thus we have excellent
agreement of both the position and the height of the peak.
The low field resistance for the  experimental $I(V)$ curves
is  $R\approx 570\Omega$ for the 35 well superlattice
(Fig. 1 of Ref.~\cite{HEL91}) and $R\approx 331\Omega$ for the 23 well 
superlattice (Fig. 2 of Ref.~\cite{HEL90}). Dividing by the number of wells
we obtain $d(eFd)/dI=16\Omega$  and  
$d(eFd)/dI=14\Omega$, respectively, which are in good agreement with
our calculated value $d(eFd)/dI=12\Omega$.
The domain branches found experimentally exhibit maximum currents
in the range  between 0.35 mA  and 0.55 mA which are lower than
$I_M$ as common for field profiles where the  
high field domain located at the cathode.

In Fig.~\ref{Figstroma} we have also plotted some data points
taken from Fig. 1 of Ref.~\cite{HEL91} (we divide the experimental voltage
by the number of periods to obtain the effective field $eFd$). 
We can see that the
onset of the second peak occurs for larger effective fields experimentally
than in our calculation.
This might be related to the fact, that a part of the voltage 
may drop outside the superlattice, so that the electric field
in the sample is smaller than the estimation $U/Nd$ used.
Good agreement is found if we assume that a voltage drop 
$U_c=0.33$ V occurs in the contact. 

\begin{figure}
\vspace*{5.1cm}
\includegraphics{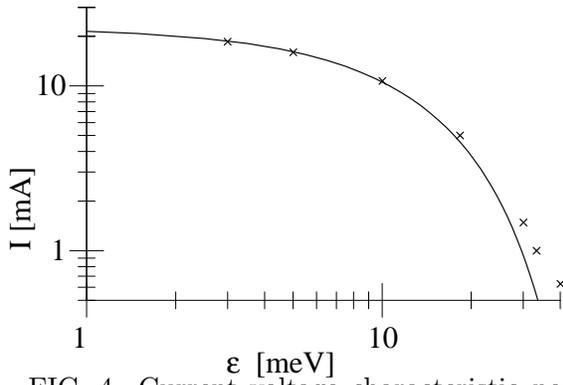}
\caption[a]{Current voltage characteristic near the second resonance.
The crosses mark experimental data taken from Ref.~\cite{HEL91}.}
\label{Figstromb}
\end{figure}

In Fig.~3 of Ref.~\cite{HEL91} the experimental current
is given in terms of $\epsilon=1.24$ meV $-U/N$. The authors do not state
that they found a maximum at $U/N =1.24$ meV but just take this value from
the measured intersubband spacing. Taking the voltage drop in the contact 
into account we have to use the value $\epsilon=124$ meV$-U_c/N-eFd$
for our theoretical data. Then we obtain
good agreement between the experimental data and the calculated
curve as shown in Fig.~\ref{Figstromb}.

\begin{figure}
\vspace*{5.1cm}
\includegraphics{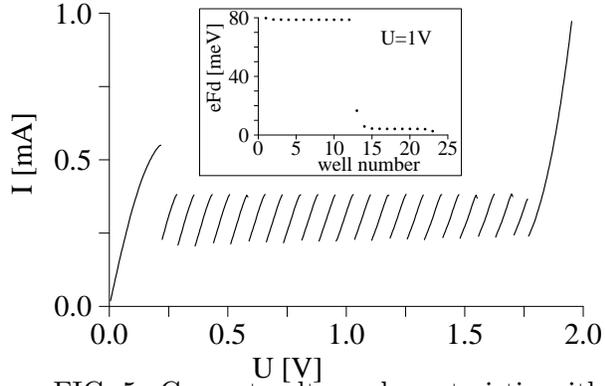}
\caption[a]{Current voltage characteristic with domain formation
for 23 wells for sweep-up of the voltage. The field distribution
for $U=1$ V is shown in the inset where the injecting contact is
located at well number 0.}
\label{Figdom}
\end{figure}

Analogously to Ref.~\cite{PRE94} we calculated the current-voltage
characteristic with domain formation using Eq.~\ref{EqJ} for the calculation
of the current from one well to the next and  the discretized Poisson
equation. Here we assumed fast relaxation within the wells, so
that the upper subband is not occupied.
Like in the experiment~\cite{HEL90} we consider a 23 well superlattice. 
For the electron density $n_i$ in well $i$ we apply the 
boundary conditions $n_0=n_{23}=0.95 N_D/A$.
The resulting current-voltage characteristic for sweep up
is shown in Fig.~\ref{Figdom} which is in good agreement with the 
experimental findings.
The high field domain is located at the anode
in our simulation which is typical for the boundary conditions 
used  and for sufficiently large  dopings as can 
easily be understood within the general theory given in
Ref.~\cite{WACp}.

\section{Conclusion}
We have developed  a microscopic model to calculate the current in 
coupled quantum wells. The self-energies, which give the renormalized 
energies and widths of the spectral functions, are calculated
directly from the nominal parameters describing the superlattice,
without invoking additional adjustable parameters.
The results for the current 
are in good quantitative agreement with the data
from Ref.~\cite{HEL90} if we assume the barrier width to be 1.5 nm less than
the nominal value. We find good agreement for the 
low field behaviour including the position and height of the first
maximum. Regarding the second peak the shape of the resonance
is in good agreement if we assume, that 0.33 V of the total voltage
drop outside the superlattice. 
Furthermore we have shown that the actual shape of the spectral functions
due to the scattering seems to be responsible for a
significant shift in the maximum of intersubband absorption.

\section{Acknowledgement}
We want to thank B. Hu and K. Johnsen for helpful discussions.
One of us (A.W.) gratefully acknowledges financial support from the German
science foundation DFG. 


\begin{table}
\begin{tabular}{|rl|}
\hline
$E^a$&$=41.4$ meV \\ 
$E^b$&$=159.3$ meV \\ 
$T_1^a$&$=-0.00539$ meV (-0.0152 meV)\\
$R_1^{ba}$&$=1.61\cdot 10^{-4}d$ $(4.36\cdot 10^{-4}d)$\\
$\Gamma_{ph}^b$&$=0.38$ meV \\
\hline
\end{tabular}
\caption[a]{Calculated parameters 
for the GaAs-Al$_{0.3}$Ga$_{0.7}$As superlattices with well width 8 nm
and barrier width 12 nm. In brackets
we have given the respective values for $b=10.5$ nm.}
\label{Thelgpar}
\end{table}

\end{document}